\documentclass[twocolumn,showpacs,aps,pra,amsmath,amssymb,superscriptaddress]{revtex4-2}
\usepackage[utf8]{inputenc}
\usepackage{newunicodechar}
\newunicodechar{⁻}{\textsuperscript{-}}

\usepackage{graphicx}
\usepackage{dcolumn}
\usepackage{bm,bbm}
\usepackage{hyperref}
\usepackage{subfigure}
\usepackage[ruled,linesnumbered]{algorithm2e}
\usepackage{soul,color,xcolor}   
\usepackage{overpic}

\usepackage{epstopdf}

\usepackage{booktabs}
\usepackage{amsmath}
\usepackage{graphicx} 

\usepackage{color,bbm}

\newcommand{\beq}{\begin{equation}}
\newcommand{\eeq}{\end{equation}}
\newcommand{\bqa}{\begin{eqnarray}}
\newcommand{\eqa}{\end{eqnarray}}
\newcommand{\nn}{\nonumber}

\definecolor{maroon}{rgb}{0.7,0,0}

\definecolor{ngreen}{rgb}{0.3,0.7,0.3}

\definecolor{golden}{rgb}{0.8,0.6,0.1}

\begin{document}

\title{Point upsampling networks for single-photon sensing}

\author{Jinyi Liu}
	\affiliation{State Key Laboratory of Autonomous Intelligent Unmanned Systems, Shanghai Research Institute for Intelligent Autonomous Systems, Tongji University, Shanghai, 201203, China}
    \affiliation{The Department of Structural Engineering, Tongji University, Shanghai, 201804, China}
    
  \author{Guoyang Zhao}
\affiliation{The Department of Control Science and Engineering, Tongji University, Shanghai 201804, China}

 \author{Lijun Liu}
  \email{lljcelia@126.com}
\affiliation{College of Mathematics and Computer Science, Shanxi Normal University, Linfen 041000, China}

\author{Yiguang Hong}
\affiliation{State Key Laboratory of Autonomous Intelligent Unmanned Systems, Shanghai Research Institute for Intelligent Autonomous Systems, Tongji University, Shanghai, 201203, China}
\affiliation{The Department of Control Science and Engineering, Tongji University, Shanghai 201804, China}

\author{Weiping Zhang}
\email{weiping\_zh@tongji.edu.cn}
\affiliation{The Department of Structural Engineering, Tongji University, Shanghai, 201804, China}

\author{Shuming Cheng}
\email{shuming\_cheng@tongji.edu.cn}
\affiliation{State Key Laboratory of Autonomous Intelligent Unmanned Systems, Shanghai Research Institute for Intelligent Autonomous Systems, Tongji University, Shanghai, 201203, China}
\affiliation{The Department of Control Science and Engineering, Tongji University, Shanghai 201804, China}

\date{\today}

\begin{abstract}
	
  Single-photon sensing has generated great interest as a prominent technique of long-distance and ultra-sensitive imaging, however, it tends to yield sparse and spatially biased point clouds, thus limiting its practical utility. In this work, we propose using point upsampling networks to increase point density and reduce spatial distortion in single-photon point cloud. Particularly, our network is built on the state space model which integrates a multi-path scanning mechanism to enrich spatial context, a bidirectional Mamba backbone to capture global geometry and local details, and an adaptive upsample shift module to correct offset-induced distortions. Extensive experiments are implemented on commonly-used datasets to confirm its high reconstruction accuracy and strong robustness to the distortion noise, and also on real-world data to demonstrate that our model is able to generate visually consistent, detail-preserving, and noise suppressed point clouds. Our work is the first to establish the upsampling framework for single-photon sensing, and hence opens a new avenue for single-photon sensing and its practical applications in the downstreaming tasks.


\end{abstract}

\maketitle

\section{Introduction}\label{Sec1: Introduction}

Single-photon sensing, enabling optical imaging at the fundamental limit of light detection~\cite{kirmani2014first, singlequantum2, singlequantum1, wang2023mid, lee2023caspi}, has recently achieved significant progress, including single-photon avalanche diode (SPAD) detectors fabricated with picosecond-level temporal resolution~\cite{chen2022deep, hadfield2009single, piron2020review, shin2015photon} and data-processing algorithms developed for ultra-sensitive imaging~\cite{spad1, spad2, de2021phonon}. As it is suitable to do the sensing task under extreme environments, such as turbid water~\cite{maccarone2015underwater}, dense fog~\cite{shi2022noise, zhang2022three}, and long-distance capture~\cite{pawlikowska2017single, li2020super, li2020single, li2021single, hadfield2023single}, it has generated great interest from academia and industry.

However, when single-photon sensing is applied to capture the target scene in practice, it still faces several issues. First, since single-photon imaging is inherently limited by detector design and photon budget~\cite{spatial1,spatial2}, low-spatial resolution images are usually generated~\cite{li2020super, li2020single, li2021single, airborne}. Then, it suffers from offset noise~\cite{kirmani2014first, shin2015photon, rapp2017few} which introduces depth bias to signal photons and also degrades geometric fidelity~\cite{sparse}. Besides, there is noise induced by dark counts of detectors and ambient lights. These limit this sensing technique in practical applications, such as fine reconstruction and semantic interpretation~\cite{object2024,task1,task2, task3, task4}.

Here, we introduce the technique of point upsampling (PU) to single-photon sensing, by using PU networks to increase spatial resolution and reduce distortion noise for single-photon point cloud. Particularly, our learning network is established on the state space model (SSM), or equivalently the Mamba architecture~\cite{Gu2023Mamba}, hence called as SPU-MAMBA. Given any single-photon point cloud, our model as shown in Fig.~\ref{upsampling} integrates a multi-path scanning mechanism to enhance spatial consistency, a bidirectional Mamba backbone to capture global context and local geometry, and an adaptive upsample shift module to mitigate point non-uniformity and depth distortion. Thus, it enables densification and correction of sparse and spatial-biased single-photon data.

Importantly, the proposed SPU-MAMBA is able to upsample non-uniformly distributed points with linear computational complexity to have a much more continuous representation of scene geometry~\cite{Zhang2024Pointmamba,Liu2024Pointmamba}. By contrast, extra assumptions are needed for other methods to do the PU task. For example, CNN-based models~\cite{Pu-net} need uniform sampling and smooth local structures, graph-based ones~\cite{PU-GCN} rely on stable local connectivity, and diffusion-based ones~\cite{MPU, Grad-PU, PU-transformer} overlook point clustering and range-dependent depth shifts.  

Extensive experiments are implemented to validate our SPU-MAMBA. First, the results obtained on commonly-used datasets confirm that our approach achieves higher reconstruction accuracy and stronger noise robustness, in comparison to other PU methods. Then, our results on real-world data demonstrate that it can generate visually consistent and detail-preserving single-photon point clouds, while effectively suppressing distortion noise. These highlight the reliability and generalizability of our approach in 3-dimensional (3D) perception at the fundamental level, hence paving the way for exploring single-photon sensing in practical applications.

This work is structured as follows. Sec~\ref{sec:problem definition} first provides an overview of single-photon point cloud in photon detection, and Sec~\ref{sec:method} then introduces the complete PU framework to address the main issues in single-photon sensing.  Further, Sec~\ref{methodology} establishes the experimental setup, and Sec~\ref{sec:results} presents our experimental results. Finally, it is concluded with further discussions in Sec~\ref{sec:discussion} and a brief summary about our work.

\section{Single-photon sensing: principles, algorithms, and point clouds}\label{sec:problem definition}

In this section, we first give a brief description of how single-photon data are generated, then recap some commonly-used algorithms to process these data, and finally introduce the single-photon point cloud.

\subsection{Photon counting}

\begin{figure}
		\begin{center}
	\includegraphics[width= 1.0\columnwidth]{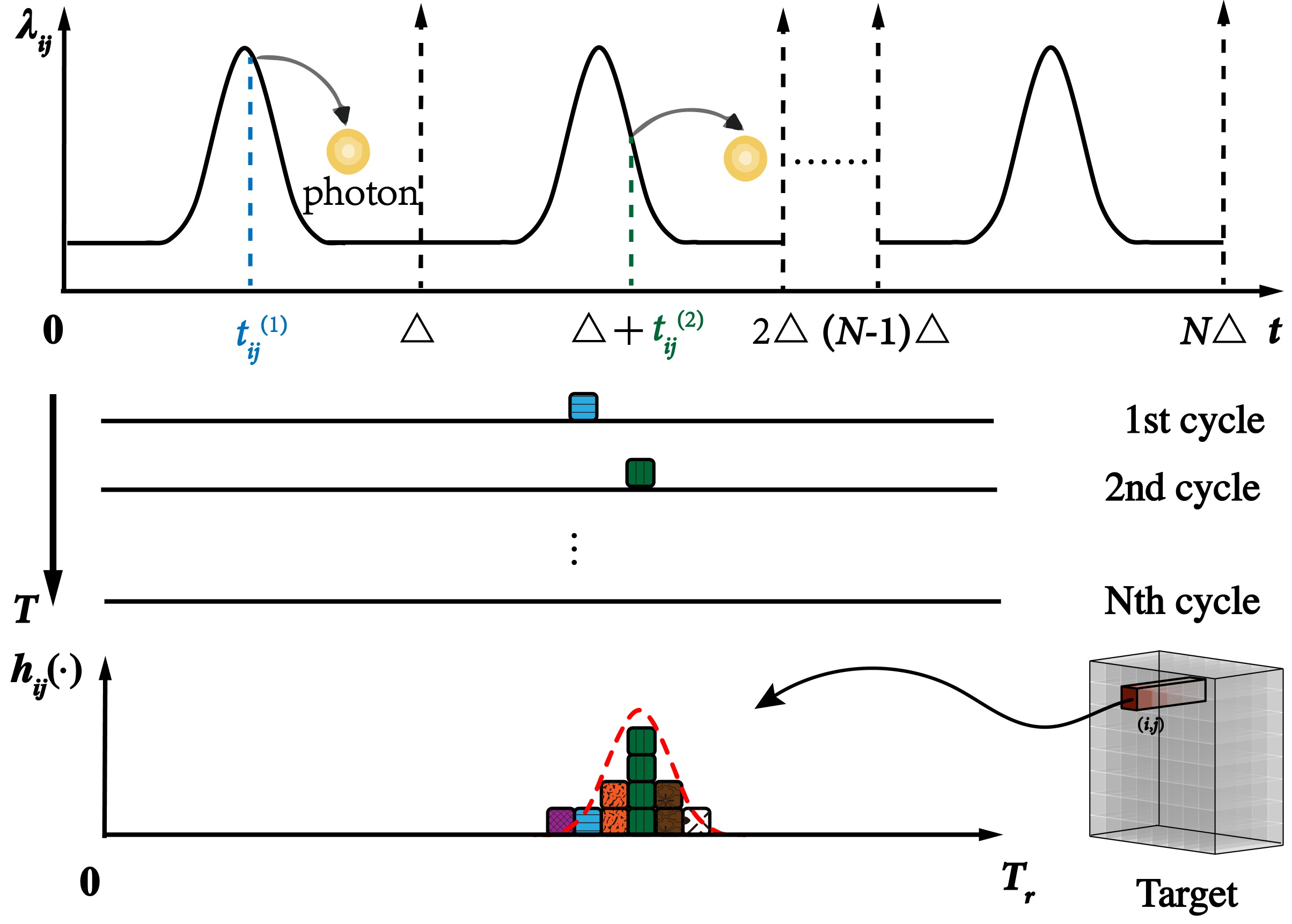}
\end{center}
	\caption{For pixel $(i, j)$, its single-photon counting histogram $h_{ij}(\cdot)$~(\ref{histogram}) follows approximately a Poisson distribution~(\ref{counts}).}
	\label{single-photon}
\end{figure}

The typical LiDAR system based on single-photon sensing contains three components: a pulsed laser source, a SPAD detector, and time-correlated single-photon counting (TCSPC) modules. It works that the laser source sends periodic pulses of light to illuminate the target scene, the SPAD detector collects reflected photons, and the TCSPC module records the time of flight (ToF) of these received photons. As illustrated in Fig.~\ref{single-photon}, the data generated from this system obeys fundamental principles. 

Each target pixel, labelled by $(i,j)$, is established from periodic light pulses over a time horizon $T$, which is equally split into $N$ intervals with time-interval $\Delta= T/N$. During each interval $n=1,\dots, N$, one single light pulse is sent out from the laser source; the SPAD records the number of reflected photons $c_{ij}[n]$, and correspondingly, the TCSPC records their ToFs as vector $\boldsymbol{t}_{ij}[n]$. Since the low-light condition is generally satisfied that at most one photon is detected within one single interval~\cite{shin2015photon}, the ToF vector is simplified to scalar $t_{ij}[n]$ and thus a pair of data $(t_{ij}[n],c_{ij}[n])$ is collected. 

In principle, $N$ pairs of $(t_{ij}[n],c_{ij}[n])$ can be converted into a photon-counting histogram, of which the time window is chosen as $[0, T_r]$ with $ T_r > \max_n t_{ij}[n]$ and $K$ ranges are sampled as $[k\delta,(k+1)\delta)$ with $k=0,\dots, K-1$ and sample precision $\delta$. Specifically, the total number of photons with which ToF falls into each sampled interval $[k\delta,(k+1)\delta)$ is counted as
 \begin{align}
 	h_{ij}[k]=\sum_{n=1,~t_{ij}[n]\in [k\delta,(k+1)\delta)}^N c_{ij}[n] \label{histogram}
 \end{align}
Further, the actual density of photons that arrive at the SPAD within each interval $n$, called photon flux, is given by~\cite{shin2017photon}
 \begin{equation}\label{foward}
 	r_{ij} [n]=\int_{n\Delta}^{(n+1)\Delta } \Phi_{ij}\cdot \alpha_{ij}\cdot s\left(t-\frac{2z_{ij}}{c}\right)dt+B_\gamma  
 \end{equation}
where $\Phi_{i j}$ represents the light attenuation, $\alpha_{ij}$ the target's reflectivity, $s(\cdot)$ the temporal wave function of light pulse, $c$ the speed of light, $B_\gamma$ the background noise at frequency $\gamma$, and $z_{i j}$ the distance between the laser source and the target. As the light pulse in each interval has an identical waveform, the photon flux~(\ref{foward}) are identically distributed, i.e., $r_{i j}[n]= r_{ij}$ for all $n$. Consequently, the histogram $h_{ij}(\cdot)$ follows approximately a Poisson distribution~\cite{shin2017photon}
 \begin{equation}
 	h_{ij}(\cdot) \sim P_{N(\eta\,r_{ij}+B_d )}  \label{counts}
 \end{equation}
 Here, $\eta \in [0,1]$ is the detector's detection efficiency and $B_{d}$ the thermally-induced dark count of the detector. 
 
After illuminating the entire scene over $i=1,\dots, I$ and $j=1,\dots, J$,  the collected data give rise to a histogram matrix
\begin{equation} 
	\mathbf{H}:=\{h_{ijk}: i=1,\ldots,I; j=1,\ldots,J; k=0,\ldots,K-1\} \label{data}
\end{equation}
where element $h_{ijk}$, equal to $h_{ij}[k]$ in Eq.~(\ref{histogram}), describes the photon counting of pixel $(i,j)$ within time bin $k$. Note that $I\times J$ represents the spatial resolution, and $K$ the temporal resolution.

\subsection{Algorithms}

To accomplish the sensing task, such as reconstruction, classification, and segmentation, various algorithms have been developed to process single-photon data encoded in histogram matrix $\mathbf{H}$.  Generally, these algorithms fall into two categories: One is model-based and the other learning-based. The first combines the photon-counting model~(\ref{foward}) with histogram data~(\ref{histogram}) to formulate an optimization problem, and notable examples include first-photon imaging~\cite{kirmani2014first}, filtering and denoising~\cite{rapp2017few}, and three-dimensional deconvolution~\cite{li2021single}. Alternatively, the latter leverages the power of machine learning to extract target features directly from collected data, and examples of this paradigm include non-local neural network~\cite{peng2020photon}, U-net++ architecture~\cite{zang2021non}, and pixel-wise residual shrinkage net~\cite{yao2022robust}.

It is remarked that the underlying algorithm used in single-photon sensing could reduce noise, extract features, and refine interpretations, however, it cannot output more information and finer details than those captured by the real hardware, which are fundamentally limited by the spatial and temporal resolutions of sensors.

\subsection{Single-photon point cloud}

\begin{figure}
	\centering
	\includegraphics[scale=0.25]{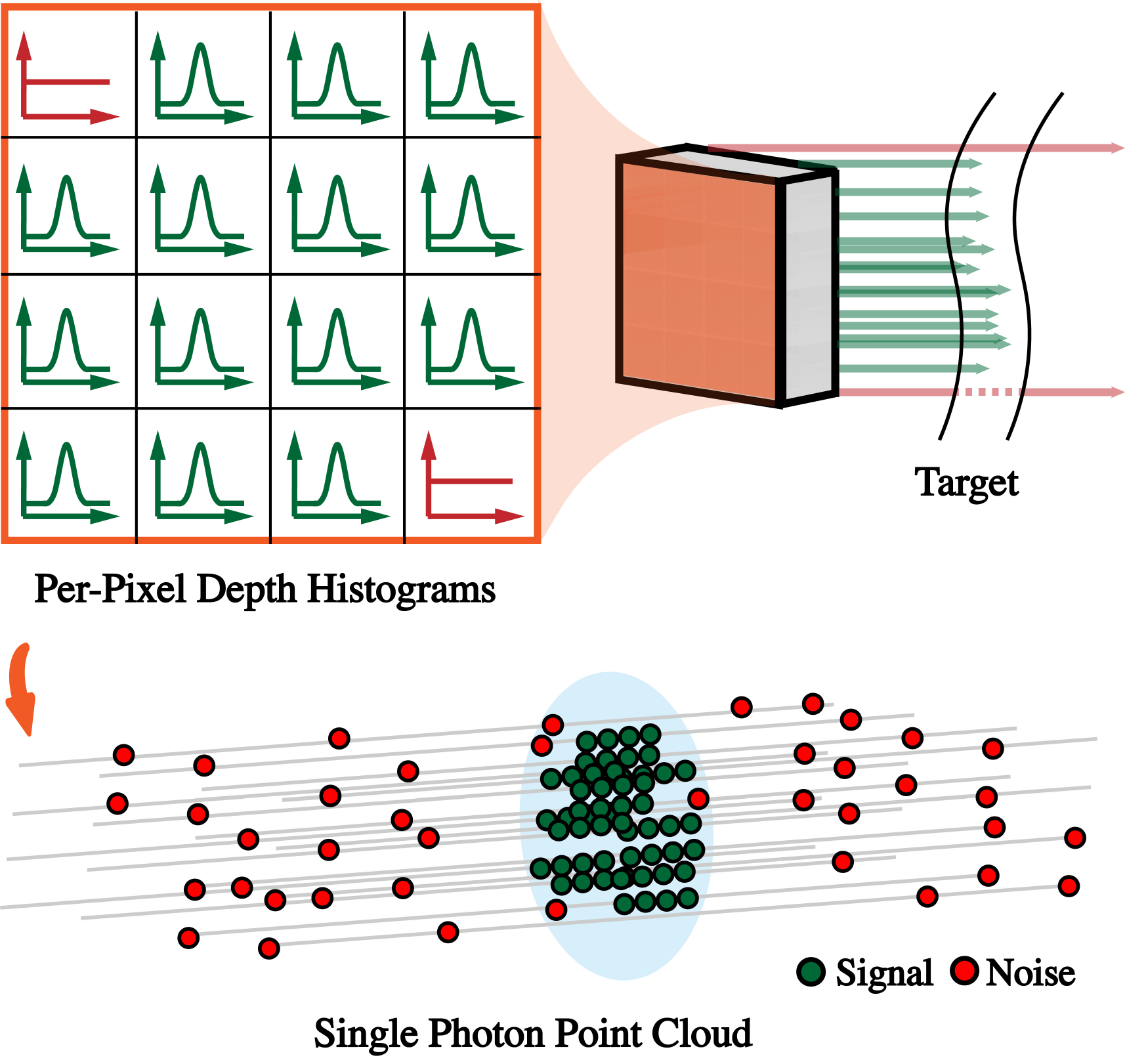}
	\caption{The photon-counting histogram data $\mathbf{H}$ in Eq.~(\ref{data}) gives rise to a single-photon point cloud $\mathcal{P}_I$~(\ref{initialpoint}), which in practice suffers from noise and limited resolution}
	\label{characteristic}
\end{figure}

The histogram matrix $\mathbf{H}$ in Eq.~(\ref{data}) can be refined into a coordinate matrix
\begin{equation} 
	\mathcal{P}_I := \{\boldsymbol{p}=(x, y, z) \mid h_{ijk} > 0\} \label{initialpoint}
\end{equation}
with
\begin{align}  \label{coordinate}
x = (i - i_c) \cdot \frac{z}{f}, ~y = (j - j_c) \cdot \frac{z}{f}, ~z = \frac{c \cdot k \cdot \delta}{2}
\end{align}
Here $(i_c, j_c)$ denotes the coordinates of principal point, $f$ the focal length, $c$ the speed of light, and $\delta$ the time-bin width. Geometrically, $\mathcal{P}_I$ gives rise to a point cloud where each point is determined by 3D coordinates  defined in Eq.~(\ref{coordinate}).

In practice, the single-photon point cloud $\mathcal{P}_I$ suffers from the following issues.

\textbf{Noise and distortion}: As shown in Fig.~\ref{characteristic}, it contains noise, such as ambient lights and dark counts of SPAD detectors~\cite{spad1,spad2}, and also depth bias, due to photon arrival statistics and incidence angle correlations~\cite{kirmani2014first,shin2015photon}. Besides, it has distortion that points follow a highly non-uniform spatial distribution governed by surface reflectivity rather than the underlying geometry~\cite{rapp2017few}, generating dense point clusters in highly reflective regions and leaving less reflective areas sparsely sampled. 

\textbf{Limited spatial resolution}: There exists a trade-off relation between depth precision and point sparsity~\cite{scholes2023fundamental}, in the sense that the finer temporal binning improves depth precision but spreads photons across more time bins to increase sparsity, and conversely, the coarser binning increases the point density but sacrifices depth resolution. As a consequence, it is produced with limited spatial resolution, failing to meet requirements of local curvature continuity and edge integrity~\cite{sparse}.

The above issues are detrimental to the performance of single-photon point cloud in the downstream tasks, such as object detection, scene segmentation, and geometric classification, thus heavily limiting the practical utility of single-photon sensing. More challenges with conventional upsampling methods for SPAD data are deferred to Appendix~\ref{Challenges}. In the following, learning-based techniques are leveraged to address these issues.

\section{Point upsampling}\label{sec:method}

In this work, we apply the technique of PU to single-photon sensing, by using PU networks to increase spatial resolution and reduce distortion noise for single-photon point cloud. Specifically, SPU-MAMBA is proposed to generate synthetic points that are expected to lie on the underlying geometric surface, thus yielding an upsampled point cloud $\mathcal{P}_U$ from an initial $\mathcal{P}_D$, with $N_U = N_D \times r_c$ where $N_U$ ($N_D$) is the number of points in $\mathcal{P}_U$ ($\mathcal{P}_D$) and $r_c$ the upsampling rate. As it is essentially built on SSM, or equivalently Mamba~\cite{Gu2023Mamba}, it can capture long-range geometric dependencies with linear computational complexity~\cite{Zhang2024Pointmamba,Liu2024Pointmamba}. Moreover, it introduces a selective scanning mechanism to dynamically deal with non-uniformly distributed points, going beyond attention-based approaches~\cite{Attention} and hence being able to generate upsampled point clouds with a much more continuous representation of scene geometry. More discussions about Mamba are deferred to Appendix~\ref{SSM}.

\begin{figure*}[htbp]
		\centering
		\includegraphics[width=\linewidth,scale=0.80]{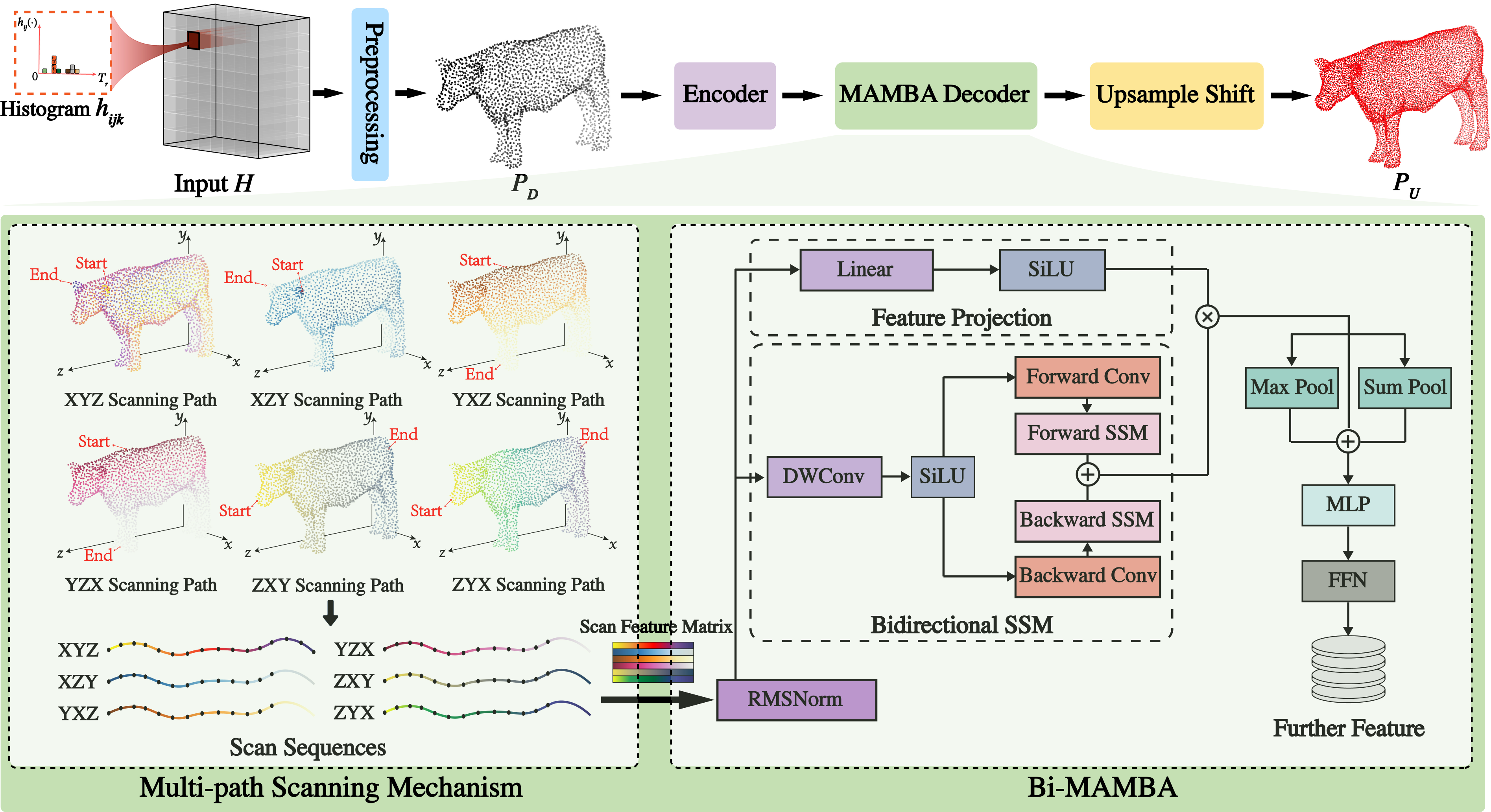}
		\caption{The SPU-MAMBA architecture contains three components: Encoder, MAMBA Decoder, and Upsample Shift. Any point cloud $\mathcal{P}_D$ is first encoded into high-dimensional feature representations, then serialized via multi-path scanning mechanism and further processed by bidirectional Mamba blocks to learn deep features from multiple perspectives, and finally upsampled into $\mathcal{P}_U$ via Upsample Shift. It is essentially built on SSM, or Mamba~\cite{Gu2023Mamba}, which can capture long-range geometric dependencies with linear computational complexity~\cite{Zhang2024Pointmamba,Liu2024Pointmamba}.}
		\label{upsampling}
	\end{figure*}

The proposed SPU-MAMBA is configured with three core components: a point encoder, a Mamba decoder, and an upsample shift module, following the cascaded structure of feature encoding, spatio-temporal decoding, and geometric transformation.  First, point cloud $\mathcal{P}_D$ is fed into the point encoder to output point-wise embeddings $F_{e}$. Then, the Mamba decoder utilizes the multi-path scanning mechanism to serialize these point embeddings, which is further processed by cascaded Bi-Mamba blocks. Subsequently, decoded features $F_d$ are concatenated with encoder features $F_e$, all of which are regulated by feed-forward network (FFN) and multi-layer perception (MLP). Finally, the upsample shift module integrates these features to predict offsets for $r_c$ generated points associated with each input point from $\mathcal{P}_D$, and applying these offsets immediately yields an upsampled point cloud $\mathcal{P}_U$, with upsampling rate $r_c$.  Its complete working pipeline is outlined in Fig.~\ref{upsampling}, and each of the network component is detailed below.

\subsection{Point Encoder} 

The point encoder transforms any point cloud $\mathcal{P}_D \in \mathbb{R} ^{N_D\times 3}$ into high-dimensional point-wise embeddings $F_{e} \in \mathbb{R} ^{N_D\times C_{e}}$, with feature dimensionality $C_e$. Especially, local geometric structures in point cloud are captured through a series of edge convolution operations that dynamically construct local neighborhood graphs for each point~\cite{wang2019dynamic}, and semantic information critical for the decoding process are also preserved in these embeddings.

\subsection{Mamba Decoder}\label{decoder}

 It is shown in the light-green box of Fig.~\ref{upsampling} that the Mamba decoder first employs a multi-path scanning mechanism to serialize point coordinates and to generate permutation-equivariant sequences that preserve three-dimensional spatial relationships. Then, these serialized sequences are processed by bidirectional Mamba blocks with $\mathcal{O}(N_D)$ computational complexity, further enhanced by recalibrating information flow between the encoded $F_{e} \in \mathbb{R} ^{N_D\times C_{e}}$ and decoded $F_d \in \mathbb{R} ^{N_D\times C_{d} }$ via FFN. Notably, this architecture has the advantages of robust feature learning under limited sampling conditions and noise propagation suppression.

\textbf{Multi-path scanning mechanism}: Six paths in total are followed to scan point cloud $\mathcal{P}_D$. For example, the XYZ scanning path is given as  
\begin{equation}\label{XYZ} 
	\mathcal{S}{xyz} (\mathcal{P}_D)= \text{Sort}_{z}\bigg(\text{Sort}_{y}\big(\text{Sort}_{x}(\mathcal{P}_D)\big)\bigg) 
\end{equation} 
where function $\text{Sort}_k(\cdot)$ sorts points in ascending order according to the coordinate value along axis $k \in {x,y,z}$. The nested structure indicates that points are first sorted along $x$ axis, then along $y$ axis with identical $x$ values, and finally, along $z$ axis with identical $x$ and $y$ values. Its complementary scanning paths, including XZY and its permutations, can be defined as Eq.~(\ref{XYZ}) similarly.

The multi-path scanning mechanism can handle point cloud in which dense- and sparse- alternating point regions are involved~\cite{Gu2023Mamba}. Moreover, it is able to capture both spatial coherence and geometric continuity of point cloud, by noting that
\begin{itemize}
    \item Depth-axis scanning (e.g., YXZ, ZXY) preserves surface-aligned photon clusters through sequential aggregation along object normals.
    \item Cross-pixel scanning (e.g., XZY, XYZ, YZX, ZYX) establishes connectivity between spatially adjacent but coordinate-distant points.
\end{itemize}

As illustrated in Fig.~\ref{upsampling}, the scanning mechanism is operated with directional tokenization to generate six context-aware sorting sequences that encode depth-layer continuity and local neighbourhood consistency, outperforming those based on Hilbert curves and octree-based serialization~\cite{Liang2024Pointmamba,Liu2024Pointmamba}.

\textbf{Bidirectional Mamba blocks}: In comparison to Mamba~\cite{Liu2024Pointmamba}, three variants are introduced to address the following issues in upsampling: unidirectional causality, insufficient local geometric sensitivity, and limited cross-scale feature interaction. 

First, it is built on a bi-directional model which contains one forward $\text{SSM}_{\text{forward}}$ and one backward $\text{SSM}_{\text{backward}}$, overcoming the unidirectional causality constraint in learning feature. Indeed, $\text{SSM}_{\text{forward}}$ processes feature embeddings in the forward direction, and $\text{SSM}_{\text{backward}}$ in the reverse direction, in the form of
\begin{equation}
\mathcal{F}_{\rightarrow}(F_{e,\,\mathcal{S}}) = \text{SSM}_{\text{forward}}\big(\text{LN}(F_{e,\,\mathcal{S}})\big)
\end{equation}
and
\begin{equation}
\mathcal{F}_{\leftarrow}(F_{e,\,\mathcal{S}}) = \text{SSM}_{\text{backward}}\big(\text{LN}\left(\text{Reverse}(F_{e,\,\mathcal{S}})\right)\big) 
\end{equation}
where $F_{e,\,\mathcal{S}}$ describes the feature embedding following a scanning path $\mathcal{S}$, and layer normalization (LN) is utilised to ensure numerical stability. 

Then, dual convolutional layers are placed prior to the above bidirectional SSM to capture neighbourhood relationships of feature representations, and further the resulting features are processed by maximum-pooling, mean-pooling, and concatenation with input features to construct global aggregated features. This tackles the problem that distant contextual features introduce redundancy and local geometric patterns are insensitive.

Finally, the FFN composed of two linear layers and one ReLU is employed as an adaptive modulation mechanism for decoded features, inspired by Transformer~\cite{Attention}. It can actively regulate information flow across different feature levels, thus enabling cross utilization of the encoder feature $F_e$ and Mamba decoder $F_d$.

\subsection{Upsample shift}

\begin{figure}
	\centering
	\includegraphics[scale=0.18]{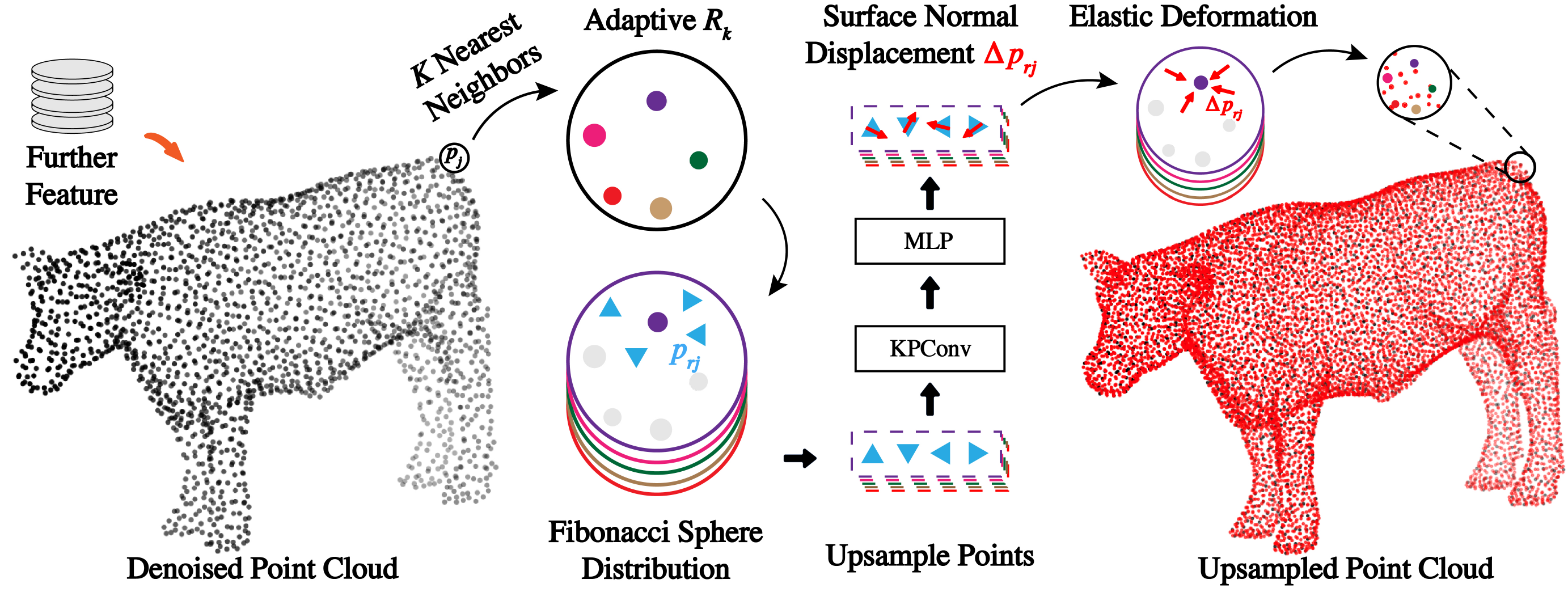}
	\caption{The upsample shift module is configured as a deformable kernel point convolution so that local geometric features of both each point and its neighbours are utilized to predict offsets for $r_c$ upsampled points.}
	\label{upsample}
\end{figure}

Since the single-photon point cloud typically follows a non-uniform spatial distribution, a two-stage deformable kernel-point convolution is configured as the shift module for upsampling, as displayed in Fig.~\ref{upsample}. For each point $\boldsymbol{p}_i$, the first stage determines its neighborhood $\mathcal{N}_K$ based on $K$ nearest points, together with a tunable radius $R_k$, and correspondingly, $r_c$ kernel points are sampled from a Fibonacci sphere distribution within this neighborhood. The second stage utilizes kernel point convolution (KPConv)~\cite{Kpconv} to extract geometric features of $\mathcal{N}_K$ and subsequently a lightweight MLP to predict displacement for $r_c$ points with the displacement vector $\Delta \boldsymbol{p}=(\Delta \boldsymbol{p}_1,\dots, \Delta \boldsymbol{p}_{r_c})$. 
Hence, the initial kernel points (blue triangles in Fig.~\ref{upsample}) are deformed by this displacement along local geometric directions, and the final positions of upsampled points for $\boldsymbol{p}_i$ are determined. The shift module is designed to generate upsampled points with a uniform distribution and sharp features preserved, while previous methods that employ a spherical uniform sampling with fixed radius~\cite{Grad-PU}  tend to create imbalances in regions with sudden curvature changes, as illustrated as the black point cloud in Fig.~\ref{upsample}.

\begin{table*}[htbp]
	\centering
	\caption{Experimental results on the PU-GAN dataset with depth offset noise $\sigma=0, 0.02, 0.06, 0.1$. The symbol $\downarrow$ means that the smaller the value is, the better performance the learning model tends to achieve. The best results are highlighted in \textbf{bold} text, while the second best are underlined.}
	\label{tab:pugan_results}
	\resizebox{\textwidth}{!}{
		\begin{tabular}{l|cccc|cccc|cccc}
			\toprule
			Method & \multicolumn{4}{c|}{CD $\times 10^{-3} \downarrow$} & \multicolumn{4}{c|}{HD $\times 10^{-3} \downarrow$} & \multicolumn{4}{c}{P2F $\times 10^{-3} \downarrow$} \\
			\cmidrule(lr){2-5} \cmidrule(lr){6-9} \cmidrule(lr){10-13}
			Noise ($\sigma$) & 0 & 0.02 & 0.06 & 0.1 & 0 & 0.02 & 0.06 & 0.1 & 0 & 0.02 & 0.06 & 0.1 \\
			\midrule
			PU-Net      & 0.735 & 1.091 & 1.246 & 1.318 & 10.764 & 17.465 & 21.085 & 21.637 & 6.737 & 11.315 & 14.963 & 16.022 \\
			MPU         & 0.632 & 0.779 & 0.965 & 1.004 & 9.453  & 10.873 & 13.003 & 12.565 & 5.125 & 8.876  & 13.082 & 13.501 \\
			PU-GAN      & 0.527 & 0.596 & 0.801 & 0.825 & 6.775  & 7.439  & 9.258  & 8.989  & 4.358 & 8.091  & 11.798 & 12.937 \\
			Dis-PU      & 0.406 & 0.728 & 0.909 & 0.930 & 5.390  & 9.041  & 11.191 & 10.566 & 3.658 & 8.911  & 12.880 & 13.870 \\
			PU-GCN      & 0.292 & 0.502 & \underline{0.688} & \underline{0.748} & \textbf{2.017}  & \textbf{4.079}  & \textbf{5.016}  & \textbf{5.297}  & 5.379 & 8.476  & 11.371 & 12.101 \\
			Grad-PU     & 0.265 & 0.555 & 0.761 & 0.781 & 2.496  & 5.241  & 6.847  & 6.890  & \underline{1.989} & \underline{6.120}  & \underline{9.181}  & \underline{9.944} \\
			RepKPU      & \underline{0.254} & \underline{0.488} & 0.708 & 0.767 & 2.473  & 5.575  & 7.967  & 7.786  & \textbf{1.889} & 6.414  & 10.450 & 11.486 \\
			\midrule
			SPU-MAMBA (Ours) & \textbf{0.248} & \textbf{0.425} & \textbf{0.639} & \textbf{0.705} & \underline{2.182}  & \underline{4.316}  & \underline{6.127}  & \underline{6.572}  & 2.824 & \textbf{5.577}  & \textbf{8.883}  & \textbf{9.807} \\
			\bottomrule
		\end{tabular}
	} 
\end{table*}

\begin{table*}[htbp]
	\centering
	\caption{Experimental results on the PU1K dataset with depth offset noise $\sigma=0, 0.02, 0.06, 0.1$. The symbol $\downarrow$ means that the smaller the value is, the better performance the learning model tends to achieve. The best results are highlighted in \textbf{bold} text, while the second best are underlined.}
	\label{tab:pu1k_results}
	\resizebox{\textwidth}{!}{
		\begin{tabular}{l|cccc|cccc|cccc}
			\toprule
			Method & \multicolumn{4}{c|}{CD $\times 10^{-3} \downarrow$} & \multicolumn{4}{c|}{HD $\times 10^{-3} \downarrow$} & \multicolumn{4}{c}{P2F $\times 10^{-3} \downarrow$} \\
			\cmidrule(lr){2-5} \cmidrule(lr){6-9} \cmidrule(lr){10-13}
			Noise ($\sigma$) & 0 & 0.02 & 0.06 & 0.1 & 0 & 0.02 & 0.06 & 0.1 & 0 & 0.02 & 0.06 & 0.1 \\
			\midrule
			PU-Net      & 0.819 & 1.138 & 1.324 & 1.458 & 11.225 & 17.703 & 19.348 & 20.464 & 2.592 & 4.946 & 6.552 & 6.651 \\
			MPU         & 0.616 & 0.892 & 1.087 & 1.179 & 8.410  & 12.286 & 14.144 & 14.961 & 1.783 & 4.038 & 5.705 & 5.663 \\
			PU-GAN      & 0.536 & 0.728 & 0.888 & 0.984 & 5.570  & 8.252  & 9.845  & 9.991  & 1.533 & 3.891 & 5.115 & 5.494 \\
			Dis-PU      & 0.533 & 0.798 & 0.998 & 1.047 & 5.499  & 9.522  & 11.192 & 11.119 & 1.556 & 3.893 & 5.373 & 5.639 \\
			PU-GCN      & 0.529 & \underline{0.676} & \textbf{0.825} & \textbf{0.872} & 3.584  & \textbf{5.164}  & \textbf{5.789}  & \textbf{6.188}  & 2.823 & 4.161 & 5.342 & 5.690 \\
			Grad-PU     & 0.548 & 0.747 & 0.959 & 1.022 & 3.374  & 6.979  & 8.265  & 8.787  & 1.460 & 2.937 & \underline{4.072} & \textbf{4.448} \\
			RepKPU      & \textbf{0.479} & 0.857 & 1.145 & 1.198 & \textbf{2.828}  & 7.184  & 9.141  & 9.479  & \textbf{0.798} & \underline{2.866} & 4.167 & \underline{4.468} \\
			\midrule
			SPU-MAMBA (Ours) & \underline{0.485} & \textbf{0.626} & \underline{0.840} & \underline{0.929} & \underline{3.331}  & \underline{5.756}  & \underline{7.244}  & \underline{7.882}  & \underline{1.160} & \textbf{2.674} & \textbf{4.031} & 4.551 \\
			\bottomrule
		\end{tabular}
	} 
\end{table*}

\subsection{Performance evaluation}

The performance of an upsampled point cloud is finally evaluated by 
\begin{equation}\label{total}
\mathcal{L} = \mathcal{L}_{cd} + \mathcal{L}_{hd} + \mathcal{L}_f + \mathcal{L}_r
\end{equation}
The first term describes the Chamfer distance between the upsampled point cloud $\mathcal{P}_U$ and ground truth $\mathcal{P}_{GT}$, which is defined as 
\begin{equation}\label{eq:cd}
\mathcal{L}_{cd} = \frac{1}{N_U}\sum_{\boldsymbol{p}\in \mathcal{P}_U}\min_{\boldsymbol{q}\in \mathcal{P}_{GT}}|\boldsymbol{p}-\boldsymbol{q}|_{2}^{2}  + \frac{1}{N_{GT}}\sum_{\boldsymbol{q}\in \mathcal{P}_{GT}}\min_{\boldsymbol{p}\in \mathcal{P}_U}|\boldsymbol{q}-\boldsymbol{p}|_{2}^{2}
\end{equation}
It quantifies global shape alignment through bidirectional nearest-neighbor matching, however, it may lead to over smoothing of sharp features, which is exemplified as ear wrinkles in Fig.~\ref{upsample}. 

The second term corresponds to the Hausdorff distance
\begin{equation}\label{eq:hd} 
\mathcal{L}_{hd} = \max \bigg(H(\mathcal{P}_U,\mathcal{P}_{GT}),H(\mathcal{P}_{GT},\mathcal{P}_U)\bigg) 
\end{equation}
with
\begin{align}
	H(\mathcal{P}_U,\mathcal{P}_{GT}) &= \mathrm{max} _{\boldsymbol{p}\in \mathcal{P}_U} \mathrm{min} _{\boldsymbol{q}\in \mathcal{P}_{GT}}  | \boldsymbol{p}-\boldsymbol{q}|_2 \nn \\ 
	H(\mathcal{P}_{GT},\mathcal{P}_U)&=\mathrm{max} _{\boldsymbol{q}\in \mathcal{P}_{GT}} \mathrm{min} _{\boldsymbol{p}\in \mathcal{P}_U} |\boldsymbol{q}-\boldsymbol{p}|_2  \nn
\end{align}
It measures the worst-case alignment error, and is used to address the issue of isolated outliers that individual distant points (e.g. blue triangles in Fig.~\ref{upsample}) might escape detection by only optimizing the average distance $\mathcal{L}_{cd}$.

The third term is the Point-to-Surface distance, or equivalently, the surface fitting loss 
\begin{equation}\label{eq:f} 
	\mathcal{L}_f = \sum_{\boldsymbol{p} \in \mathcal{P}_D} \sum_{k=1}^{r_c} \min_{\boldsymbol{p}^\prime \in \mathcal{N}(\boldsymbol{p})}  \frac{| \boldsymbol{p}^\prime - (\boldsymbol{p}_k + \Delta \boldsymbol{p}_k)|_2^2}{\sigma^2}
\end{equation} 
where point $\boldsymbol{p}^\prime$ is in the neighborhood $\mathcal{N}$ of point $\boldsymbol{p} \in \mathcal{P}_D$, $\boldsymbol{p}_k$ denotes the $k$-th kernel point, $\Delta \boldsymbol{p}_k$ corresponds to the learned displacement for $\boldsymbol{p}_k$, and $\sigma$ is the distance scaling factor. It is employed to ensure proximity between deformed kernel points and their nearest input and hence to suppress kernel drift.  

The last term describes the repulsion loss
\begin{equation}\label{eq:r}
	 \mathcal{L}_r = \sum_{\boldsymbol{p} \in \mathcal{P}_D} \sum_{k\neq k^\prime}^{r_c}q^2(\boldsymbol{p}_k + \Delta \boldsymbol{p}_k,\, \boldsymbol{p}_{k^\prime} + \Delta \boldsymbol{p}_{k^\prime})
 \end{equation} 
with overlap function $q(\boldsymbol{a}, \boldsymbol{b}) = \max\left( 0,\, 1-|\boldsymbol{a} - \boldsymbol{b}|_2/\sigma\right)$. It follows that if the distance between two deformed kernel points $\boldsymbol{p}_k + \Delta \boldsymbol{p}_k$ and $ \boldsymbol{p}_{k^\prime} + \Delta \boldsymbol{p}_{k^\prime}$ associated with the same point $\boldsymbol{p}$ is smaller than $\sigma$, then the function $q$ outputs a positive value, and consequently, squaring it penalizes overlapping kernel regions so that upsampled points are distributed uniformly in $\mathcal{P}_U$ to cover diverse geometric patterns without redundancy.

\section{Experimental methodology} \label{methodology}

\begin{figure*}[htbp]
	\centering
	\includegraphics[width=\linewidth,scale=0.70]{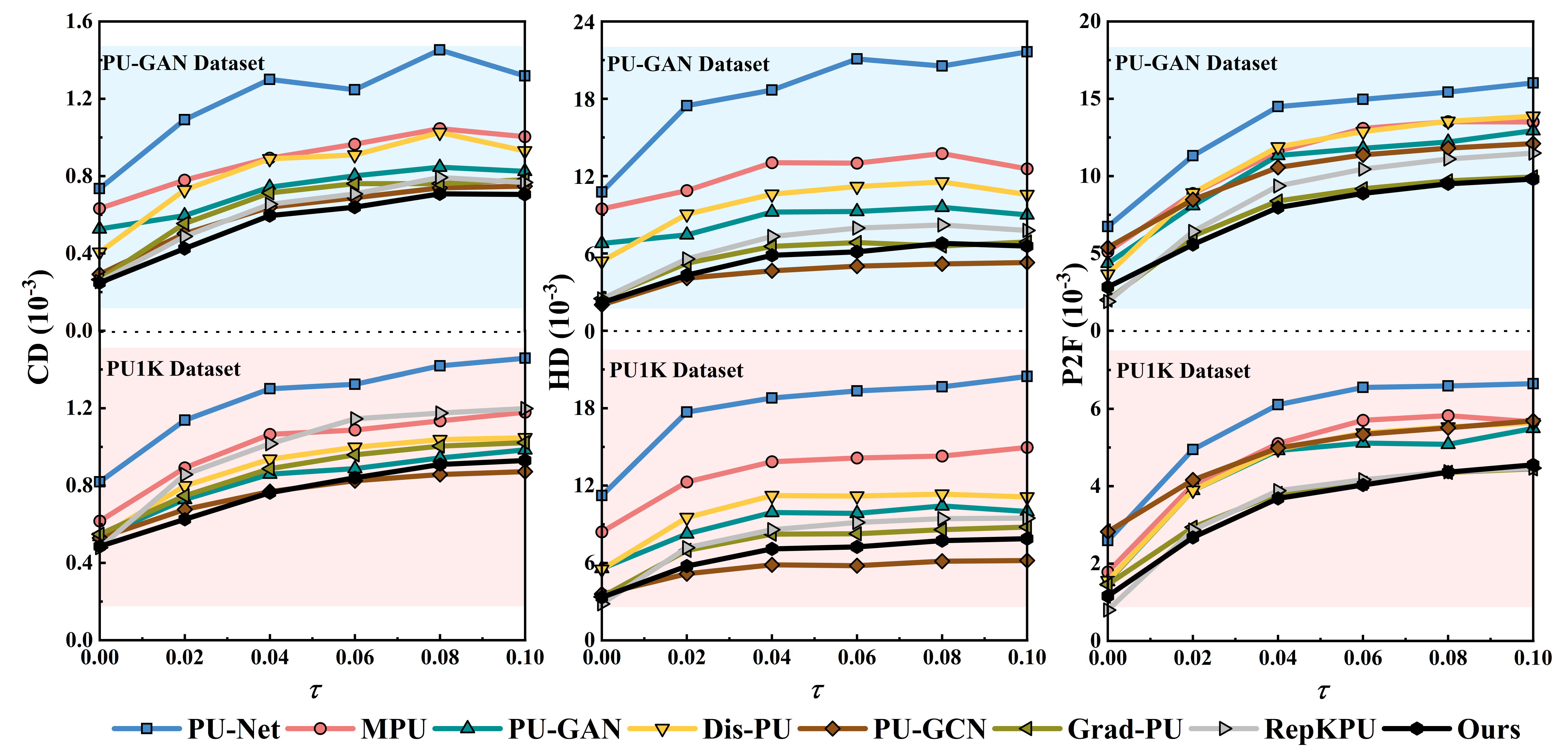}
	\caption{SPU-MAMBA is robust to the distortion noise, by achieving a flatter performance curve than other methods.}
	\label{Noise}
\end{figure*}

Two representative datasets are used to benchmark our upsampling model. One is the PU-GAN dataset~\cite{li2019pugan} which covers diverse 3D objects, including $120$ models for training and $27$ models for testing. The other is PU1K~\cite{PU-GCN} which not only comprises the complete PU-GAN, but also contains $900$ training samples and $100$ testing samples from ShapeNetCore~\cite{chang2015shapenet}. Additionally, real-world data are also collected from the single-photon LiDAR system to test the upsampling performance.

To simulate the distortion noise, a random depth offset is then added to every point in the dataset to account for the correlation between photon arrival time and emission angle, and the atmospheric scattering effect. Specifically, a depth offset $\Delta d_i$ is applied to each point with depth $d_i$, by sampling a time offset $\Delta t$ from a Gaussian distribution $\mathcal{N}(0, \sigma^2)$ with $\sigma\in [0, 0.1]$, as the depth is determined by arrival time $t$, with $d_i =c \cdot t/2$.

 Further, the state-of-the-art PU methods, including PU-Net~\cite{Pu-net}, MPU~\cite{MPU}, PU-GAN~\cite{li2019pugan}, Dis-PU~\cite{Dis-PU}, PU-GCN~\cite{PU-GCN}, Grad-PU~\cite{Grad-PU}, and RepKPU~\cite{RepKPU}, are introduced as baselines. On noise-augmented PU-GAN and PU1K, these models are trained with the settings defaulted in their original works, whereas SPU-MAMBA is with $100$ epochs and batch size $64$, and optimized via Adam with learning rate $0.001$ and beta $0.9$.

  Finally, all point clouds, including the input and its ground truth, are generated from datasets via Poisson disk sampling. For training, $50$ patches are cropped from 3D models, each of which contains $256$ points as input and $1,024$ points as ground truth. For testing, each input point cloud has  $2,048$ points and the ground truth $8,096$ points. The overlapping patches are generated by using the farthest point sampling to ensure complete coverage.

\section{Experimental results}\label{sec:results}

Extensive experiments are implemented to validate the proposed SPU-MAMBA. The upsampling rate $r_c=4$ is maintained throughout our experiments, and especially, the Chamfer distance (CD)~(\ref{eq:cd}), Hausdorff distance (HD)~(\ref{eq:hd}), and Point-to-Surface distance (P2F)~(\ref{eq:f}) are chosen as evaluation metrics.

\subsection{The upsampling performance of SPU-MAMBA}

Our experimental results on the PU-GAN dataset are listed in Tab.~\ref{tab:pugan_results} that SPU-MAMBA always scores the best CD and ranks the second in terms of HD, across all noise levels. Remarkably, it still attains the best P2F with moderate noise, outperforming the second best Grad-PU by 8.9\% at $\sigma=0.02$ and 3.2\% at $\sigma=0.06$.

The results on the PU1K dataset are listed in Tab.~\ref{tab:pu1k_results} that SPU-MAMBA achieves the best CD at $\sigma=0.02$ and ranks the second at others with minor differences. Indeed, its CD scores at $\sigma=0.06$ and $\sigma=0.1$ are only 1.8\% and 6.5\% higher than those of the best PU-GCN. Again, its HD values are consistently the second best, while its P2F performance remains at the top level, with two lowest scores and one second-lowest.

It follows immediately that our model not only admits higher reconstruction accuracy under noise, but also maintains more stable performance across dataset. This is mainly due to the multi-path scanning mechanism introduced in Sec.~\ref{decoder} which is able to learn spatial correlations between depth and pixel coordinates, and hence to enhance both detail retention and distribution regularity in upsampled outputs.

\begin{figure*}[htbp]
	\centering
	\includegraphics[width=\linewidth,scale=1.00]{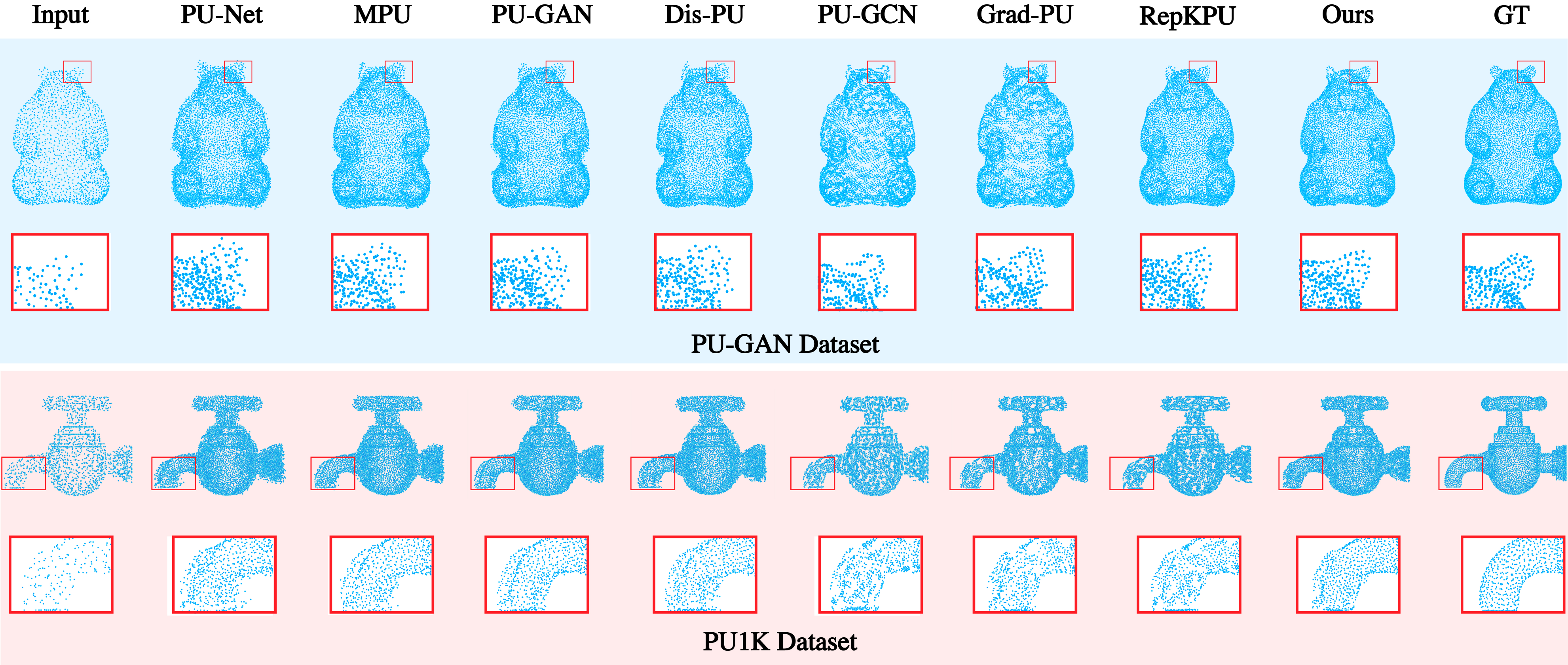}
	\caption{Visualization of the upsampling rate $r_c=4$ on PU-GAN and PU1K. SPU-MAMBA outputs a point cloud in which upsampled points are well-aligned to the ground truth.}
	\label{pu1k/pugan}
\end{figure*}
\begin{figure*}[htbp]
	\centering
	\includegraphics[width=\linewidth,scale=1.00]{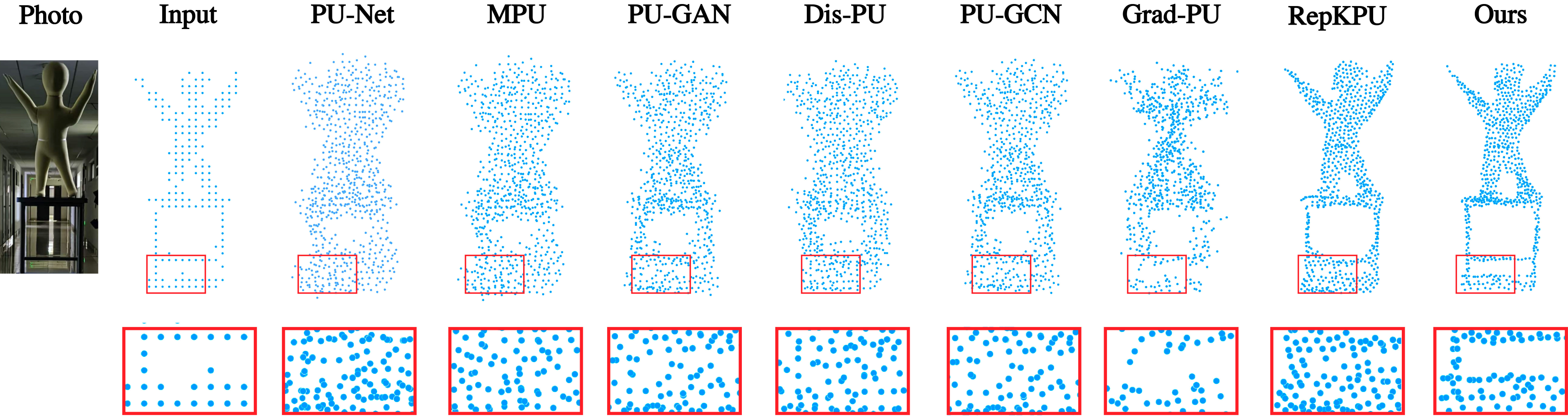}
	\caption{Visualization of the upsampling rate $r_c=4$ on real-world data. Our approach generates upsampled points that follow a more uniform distribution and have a better preservation of fine geometric details, in comparison to the baseline methods.}
	\label{realworld}
\end{figure*}

\subsection{Robustness to the distortion noise}

 The upsampling performance of these learning models with the depth-offset variance $\sigma$ varying from $0$ to $0.1$ is displayed in Fig.~\ref{Noise}. It is evident that noise degrades all methods, however, SPU-MAMBA exhibits the most consistent performance in CD and P2F, and maintains HD with minor fluctuation. Its robustness to the distortion noise is largely driven by the bidirectional Mamba decoder which captures long-range geometric dependencies and mitigates noise propagation, and the upsample shift module that dynamically adapts to non-uniform point distributions. It is also noted that PU-GCN shows a more stable HD, which can be attributed to its graph convolutional structure on the local topological consistency.

\subsection{Visual comparison on datasets with noise}

Recall from Sec.~\ref{methodology} that a random depth offset is added on the PU-GAN and PU1K datasets, so all input point clouds are mildly deviated from ground truth, as shown in Fig.~~\ref{pu1k/pugan}. Notably, SPU-MAMBA achieves a better balance to accomplish the PU task with $r_c=4$, in the sense that upsampled points are uniformly distributed and geometric sharpness is also preserved. 

By contrast, both PU-Net and MPU tends to oversmooth fine structures, as illustrated by softened contours of bear ears in PU-GAN and thickened pipe edges in PU1K.  PU-GAN and Dis-PU produce relatively uniformly distributed points, but lack geometric details, possibly leading to blob-like or overly rounded features. Although PU-GCN and Grad-PU perform well in preserving global geometry, their outputs usually exhibit the non-uniform point density so that local clustering is generated in some smooth regions. RepKPU performs competitively, but minor density fluctuations arise from the initial offset noise.

\begin{figure*}[htbp]
	\centering
	\includegraphics[scale=0.33]{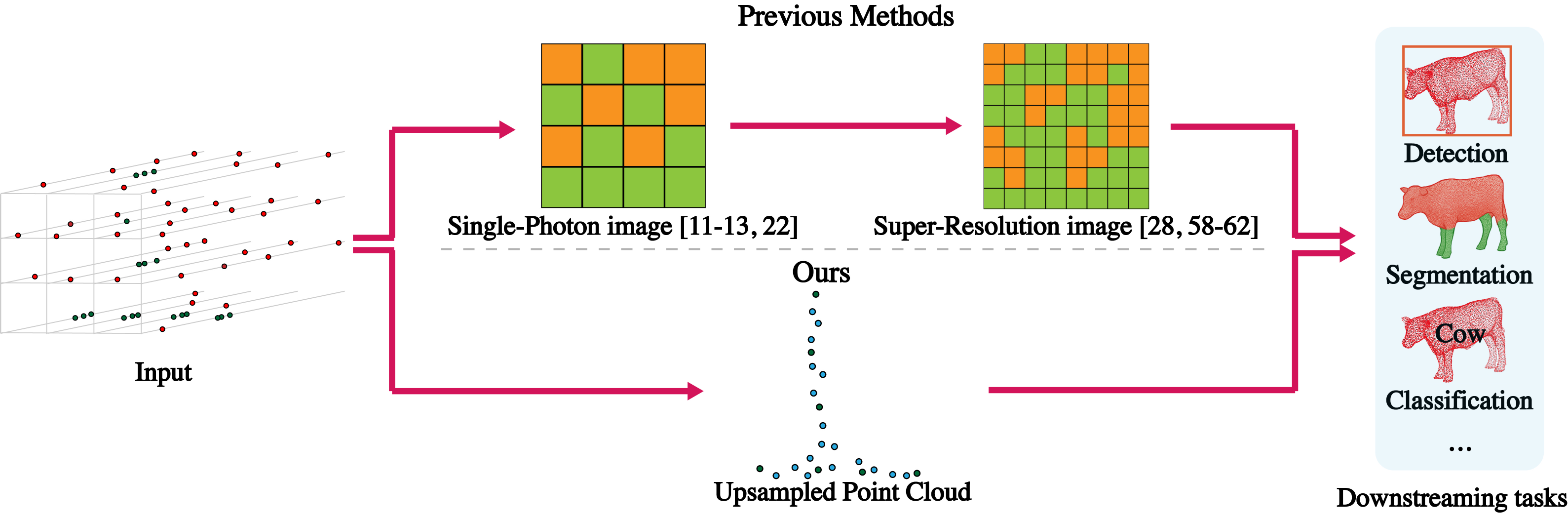}
	\caption{The comparison between PU and image super-resolution. Given raw single-photon data, PU directly increases point density and reduces noise, while the latter increases image resolution and reduces noise of single-photon images. It is highly possible that noise is imprinted as inherent details in single-photon images for super-resolution.}
	\label{denoising}
\end{figure*}

\subsection{Visual comparison on real-world data}

Our method is then evaluated on the real-world single-photon data, and the corresponding result is shown in Fig.~\ref{realworld}. Since there is no ground truth, its performance is assessed based on uniformity and detail preservation. It is found that our model is able to generate points which are visually better to distinguish human pose and greater resilience to residual noise and outliers (e.g., noise highlighted in the red box is eliminated).  However, all baseline methods exhibit some certain degradation. For example, PU-Net and MPU yield uneven surfaces and fragmented reconstructions. PU-GAN and Dis-PU admit shape flattening in thin or curved regions, while PU-GCN suffers from clustered outliers near boundaries.

\subsection{Ablation study}

The ablation study is finally conducted to validate the main components of our PU architecture for single-photon point cloud, and all experiments are implemented with $r_c=4$ on the PU1K dataset.

The effectiveness of the bidirectional Mamba decoder in SPU-MAMBA is verified by replacing it with no or unidirectional Mamba. It is obvious from Table~\ref{AblationPU1K} that our bidirectional Mamba achieves the best performance, while removing Mamba or using unidirectional Mamba degrades its performance, confirming its crucial role in providing interaction information for PU. 
\begin{table}[h!]
    \centering
    \scriptsize 
    \renewcommand{\arraystretch}{1.3}
    \caption{Experimental results with different Mamba architectures.}
    \label{AblationPU1K}
    \begin{tabular}{c|ccc}
        \toprule
        \textbf{Type of Mamba} & \textbf{CD  $\downarrow$} & \textbf{HD  $\downarrow$} & \textbf{P2F $\downarrow$} \\
        \midrule
        No Mamba & 0.606 & 4.474 & 1.508 \\
        Unidirectional Mamba & 0.546 & 3.705 & 1.385 \\
        Bidirectional Mamba (Ours) & \textbf{0.485} & \textbf{3.331} & \textbf{1.160} \\
        \bottomrule
    \end{tabular}
\end{table}

The multi-path scanning mechanism employed in SPU-MAMBA is also compared to other strategies, including random input, XYZ scanning, and space filling curves (Hilbert and Z-order)~\cite{Liu2024Pointmamba}. It is obtained in Table~\ref{AblationMamba} that our scanning strategy again score the best results, confirming that the corresponding serialization reveals spatial relationships from multiple perspectives and further enables Mamba to have a better understanding of  point interactions and to capture local geometric features. 

\begin{table}[h!]
    \centering
    \scriptsize 
    \renewcommand{\arraystretch}{1.3}
    \caption{Experimental results with different scanning mechanisms. }
    \label{AblationMamba}
    \begin{tabular}{c|ccc}
        \toprule
        \textbf{Scanning mechanism} & \textbf{CD  $\downarrow$} & \textbf{HD  $\downarrow$} & \textbf{P2F $\downarrow$} \\
        \midrule
        Random input & 0.583 & 3.775 & 1.368 \\
        XYZ Scanning & 0.541 & 3.561 & 1.323 \\
        Hilbert Scanning  & 0.518 & 3.530 & 1.276 \\
        Z-order Scanning  & 0.530 & 3.542 & 1.318 \\
        Multi-path Scanning (Ours) & \textbf{0.485} & \textbf{3.331} & \textbf{1.160} \\
        \bottomrule
    \end{tabular}
\end{table}

\section{Discussions}\label{sec:discussion}

This work presents a PU network to increase point density and also to reduce distortion noise for single-photon point cloud, and correspondingly, the upsampled point cloud is used to do the downstream tasks, such as detection, segmentation, and classification. Alternately, raw single-photon data could be first processed as images which are then improved to high-resolution ones via image super-resolution~\cite{high1, high2, high3, high4, high5, high6} and hence to do the similar task. However, the latter routine might imprint point non-uniformality and noise as inherent details into images so that they cannot be effectively reduced. These two strategies are contrasted in Fig.~\ref{denoising}.

In this work, the proposed model is only subjected to the offset-induced noise, as we the first to study PU in single-photon sensing and thus those baseline methods are not targeted for single-photon data. Furthermore, it is consistent with the prevailing practice in the relevant literatures~\cite{li2020super, li2020single, li2021single, airborne} where the geometry-based processing techniques are used under some certain noise.

It is remarked that the denoising module can be added to reduce noise before upsampling, which is useful to improve the PU performance and also to better accomplish the downstream tasks. This is left in the future work to integrate denoising functionality into our upsampling framework.

\section{Conclusions} \label{Conclusions}

We have presented a PU network based on Mamba to efficiently generate a dense and accurate point cloud from sparse and spatial-distortion sing-photon point cloud. We have also implemented extensive experiments on both synthetic and real-world data to demonstrate that our model achieves high reconstruction accuracy and strong noise robustness. Our results highlight the reliability and generalizability of our approach in 3-dimensional (3D) perception at the fundamental limit of light detection, and hence paves the way for pushing single-photon sensing into practical applications. 

There are many interesting problems to be investigated in the future. For example, it is important to extend our PU framework into a modular architecture that unifies denoising and upsampling. It is also of practice interest to consider other noise, such as background noises. And it deserves further exploration in its deployment in 3D vision applications such as robotics, autonomous driving, and remote sensing.

\acknowledgements This work is funded by the National Natural Science Foundation of China (U22B201225), the Aeronautical Science Foundation of China (2022Z073038001) and the Fundamental Research Funds for the Central Universities (22120250048).

\appendix

\section{Challenges with Conventional Upsampling Methods for SPAD Data} \label{Challenges}

This appendix explains why generic point-cloud upsampling methods are not directly applicable to SPAD point clouds. We group representative approaches and highlight SPAD-specific failure modes.

Methods such as PU-Net~\cite{Pu-net} and classical resampling and optimization works~\cite{Alexa2003, Lipman2007, Huang2013, Wu2015} implicitly assume near-uniform sampling and a correlation between point density and geometric complexity. In SPAD data, however, density primarily follows reflectivity and acquisition conditions rather than curvature: bright and retroreflective regions receive disproportionately more photons, while dark or distant areas remain sparse. As a result, CNN-based pipelines tend to oversample already-dense regions and generate low-quality points in photon-starved areas, with limited ability to correct depth-offset noise.

Graph methods (e.g., PU-GCN~\cite{PU-GCN}) are more flexible but rely on stable local neighborhoods learned from conventional LiDAR distributions. SPAD point clouds exhibit abrupt density gradients and outliers; local graph construction is easily perturbed, and these methods lack mechanisms for compensating \emph{systematic depth biases} inherent to single-photon timing.

Implicit representations and diffusion-based methods~\cite{MPU, Chen2022, Grad-PU, PU-flow} can synthesize fine details, yet they are typically trained on data that do not capture SPAD-specific effects (reflectivity-driven sparsity, depth-offset distortions). Consequently, they may interpret photon-induced artifacts as geometry, struggle to disentangle physical distortions from true structure, and hallucinate details in severely sparse regions.

Attention-based methods (e.g., PU-Transformer~\cite{PU-transformer}) establish long-range correspondences but incur quadratic complexity~\cite{Attention}, limiting scalability on large point sets. In SPAD scenes with highly non-diffuse returns, cross-region propagation can overemphasize specular responses, while the lack of explicit offset-correction mechanisms leads to suboptimal performance under reflectivity variations.

SPAD upsampling requires (i) modeling long-range geometric dependencies with \emph{linear} complexity, (ii) robustness to non-uniform, reflectivity-driven sampling, and (iii) explicit accommodation of depth-offset distortions—requirements that motivate a state-space approach.

\section{State Space Models and Mamba}\label{SSM}

This appendix discusses Mamba, a state-space model for efficient point cloud processing. The appendix covers the model formulation, discretization technique, and how selective scanning optimizes sparse and noisy data handling.

\subsection{State Space Models for Point Cloud Processing}

The challenge of processing SPAD point clouds lies in balancing long-range spatial dependencies with computational efficiency. Vanishing gradients in RNN when handling long sequences~\cite{RNNs} and the quadratic time and memory costs in transformers~\cite{Attention} highlight the need for a more efficient alternative. State-space models offer a solution~\cite{Gu2023Mamba}, utilizing a continuous latent state to capture dependencies along a serialized scan of points. This approach provides an efficient method for addressing the issues of sparse, non-uniform point clouds.

In modern control theory, SSM describe systems~\cite{Visionmamba1, Visionmamba2,Visionmamba3} by mapping input signals $x\left(t\right)\in\mathbb{R}^L$ (in our case, features from points along a scan path in $\mathcal{P}_D$) to output responses $y\left(t\right)\in\mathbb{R}^L$ via an implicit latent state $h\left(t\right)\in\mathbb{R}^{N}$. The model operates based on two core components: the state equation and the observation equation. Formally, an SSM is given as

\begin{equation}\label{eq:ssm_state} 
\dot{h}\left( t \right) =\boldsymbol{A}h\left( t \right) +\boldsymbol{B}x\left( t \right) 
\end{equation} 
\begin{equation}\label{eq:ssm_output} 
  y\left( t \right) =\boldsymbol{C}h\left( t \right)
\end{equation} 
where $\boldsymbol{A}\in \mathbb{R}^{N\times N}$ governs state evolution, $\boldsymbol{B}\in \mathbb{R}^{N\times L}$ maps inputs to state changes, and $\boldsymbol{C}\in \mathbb{R}^{L\times N}$ projects the state to outputs, all for a state size $N$.

\subsection{Discretization for Point Sequence Processing}

Discrete features along scan paths are processed by applying a zero-order hold (ZOH) discretization method~\cite{Gu2023Mamba} with a time step $\varDelta$. The discretization transforms the continuous parameters $\boldsymbol{A}$ and $\boldsymbol{B}$ into discrete counterparts $\boldsymbol{\bar{A}}$ and $\boldsymbol{\bar{B}}$:

\begin{equation}\label{eq:discrete_state} 
  h_t=\boldsymbol{\bar{A}}h{t-1}+\boldsymbol{\bar{B}}x_t 
\end{equation} 

\begin{equation}\label{eq:discrete_output} 
  y_t=\boldsymbol{C}h_t 
\end{equation} 

\begin{equation}\label{eq:discrete_A} 
\boldsymbol{\bar{A}}=\exp \left( \boldsymbol{A\varDelta } \right)
\end{equation} 

\begin{equation}\label{eq:discrete_B} 
\boldsymbol{\bar{B}}=\left( \boldsymbol{A\varDelta } \right) ^{-1}\left( \exp \left( \boldsymbol{A\varDelta }-\boldsymbol{I} \right) \right) \boldsymbol{\varDelta B} 
\end{equation} 
where $h_t$ represents the state at position $t$ along the scan path, $x_t$ is the input feature at that position, and $y_t$ is the corresponding output feature that captures contextual information from previous points.

\subsection{Mamba's Selective Mechanism for Point Clouds}

Mamba introduces a selective scanning mechanism that enhances traditional SSM~\cite{Zhang2024Pointmamba, Liu2024Pointmamba}. This mechanism adapts to the local context of point clouds by modulating transitions and readouts based on input features. The mechanism efficiently handles the non-uniform density and reflectivity variations typically found in SPAD point clouds. By emphasizing informative regions, Mamba suppresses the propagation of artifacts from sparse or noisy measurements.

Our design employs a bidirectional Mamba backbone that processes each scan path in both directions, capturing long- and short-range geometric dependencies symmetrically. This architecture achieves a balance between computational efficiency and the ability to manage the complexities of SPAD data. The model preserves geometric sharpness despite non-uniform sampling, avoids the quadratic computational cost of transformer attention~\cite{Attention}, and extends the contextual coverage compared to conventional RNN~\cite{RNNs}.

The Mamba framework addresses (i) linear time complexity for efficient processing, (ii) robustness to non-uniform and reflectivity-driven sampling,(iii) adaptability to local point distributions for better depth-offset distortion handling.

\end{document}